\documentclass[prb,aps,twocolumn,showpacs,floatfix]{revtex4}
\usepackage{graphicx}

\begin{document} 
\title
{Temperature-dependent ``phason'' elasticity in a
random tiling quasicrystal}

\author{M. Mihalkovi\v{c}~\protect\cite{MM-addr} and C. L. Henley}
\affiliation{Dept. of Physics, Cornell University, Ithaca NY 14853-2501}

\def\ss #1 {\vskip 0.3cm {\noindent \large \bf{#1}}\\}
\newcommand{\Wperp}{{W^\perp}}
\newcommand{\aR}{{a_R}}
\newcommand{\Corr}{{\Gamma}}
\newcommand{\CorrH}{{\Gamma^{\rm H}}}
\newcommand{\CorrBragg}[1]{{\Corr^{\rm Bragg}_{#1}}}
\newcommand{\Corrnode}{{\Gamma^{\rm node}}}
\newcommand{\CCK}{{\bf C}}
\newcommand{\CK}{{C}}
\newcommand{\nstar}{{n_\kappa}}
\newcommand{\mm}{{\bf m}}
\newcommand{\qq}{{\bf q}}
\newcommand{\rr}{{\bf r}}
\newcommand{\xx}{{\bf x}}
\newcommand{\xpar}{{{\bf x}^\parallel}}
\newcommand{\xperp}{{{\bf x}^\perp}}
\newcommand{\Gperp}{{{\bf G}^\perp}}
\newcommand{\Gpar}{{{\bf G}^\parallel}}
\newcommand{\epar}{{{\bf e}^\parallel}}
\newcommand{\eperp}{{{\bf e}^\perp}}
\newcommand{\vsix} {{v_{\rm 6D}}}
\newcommand{\DD}{{\bf D}}
\newcommand{\fico}{f}    
\newcommand{\cfit}{c_1}    
\newcommand{\rhomagic}{{\rho^*}}
\newcommand{\zetamagic}{{\zeta^*}}
\newcommand{\rhoo}{{\rho_0}}
\newcommand{\trhoo}{{\tilde{\rho}_0}}

\newcommand{\hh}{{\bf h}}
\newcommand{\HH}{{\bf H}}
\newcommand{\hhbar}{{\overline{\bf h}}}
\newcommand{\thh}{{\tilde \hh}}
\newcommand{\thvert}{{\tilde \HH}}
\newcommand{\thnode}{{\tilde \HH}^{\rm node}}
\newcommand{\node}{^{\rm node}}
\newcommand{\Aback}{{\bf A}_{\rm back}}
\newcommand{\hvert} {{\hh^{\rm vert}}}
\newcommand{\Ao}{\mbox{\AA}}
\newcommand{\asb}{{\alpha_{\rm short}}}
\newcommand{\Zed}{\mathbold{Z}}    
\def\PRLsection #1 {\section{#1}}     

\begin{abstract}
Both ``phason'' elastic constants have been measured 
from Monte Carlo simulations of a random-tiling icosahedral 
quasicrystal model with a Hamiltonian. 
The low-temperature limit approximates the ``canonical-cell''
tiling used to describe several real quasicrystals. 
The elastic constant $K_2$ changes sign from positive to negative 
with decreasing temperature; 
in the ``canonical-cell'' limit, $K_2/K_1$ appears to approach 
$-0.7$, about the critical value 
for a phason-mode modulation instability. 
We compare to the experiments on $i$-AlPdMn and $i$-AlCuFe.
\end{abstract}

\pacs{
61.44.Br, 
62.20.Dc,  
5.10.Ln, 
64.60.Cn}   

\maketitle

Quasicrystals possess translational long-range order, 
as manifested by 
resolution-limited Bragg peaks~\cite{niceqxtal}, 
yet possess symmetries (e.g. icosahedral) incompatible with periodicity. 
The best-understood quasicrystals can be represented as
rigid tilings (or cluster networks)
decorated by atoms in a uniform fashion~\cite{Hen91ART,Hen91CCT,Wind94,Mi96a}.
Their long-range order can then be investigated in a purely tiling framework, 
being determined by the 
``tile Hamiltonian''\cite {Hen91ART,Mi96a}),
which assigns to every tiling the corresponding structural energy. 

Despite years of study, there remain two
plausible, competing scenarios that predict quasicrystal 
long-range order (in the sense that {\it both} predict Bragg peaks in
dimension $d>2$.)
The ``ideal tiling'' scenario postulates 
that the tile Hamiltonian 
enforces an essentially unique, perfectly quasiperiodic 
ground state analogous to a perfect crystal
(just as matching rules or covering rules~\cite{Matching,Jeong94}
enforce a Penrose tiling).
Alternatively,  the (equilibrium) ``random tiling'' scenario supposes 
an ensemble of many {\it nearly} degenerate packings of the structural units.
Then the true ground state is normally a coexistence of crystal phases, 
but at higher temperatures the random-tiling quasicrystal
becomes thermodynamically stable owing to its larger entropy
(See references cited in \onlinecite{Hen91ART}). 

The two scenarios 
may be distinguished experimentally
by their diffuse scattering.  
In the random-tiling case,  every Bragg peak is surrounded
by wings of diffuse scattering with distinctive shapes predicted by
the ``phason'' elastic theory~\cite{diffuse91}. 
(See (\ref{eq-diffuse}), below.)
Those shapes were observed in X-ray and neutron diffraction
of $i$-AlPdMn\cite{Bois95,Le01} and $i$-AlCuFe\cite{Le99}, 
and the fitted phason-elastic constants were temperature-dependent.

In this Letter, we simulate random
tilings of rhombohedra with a tile Hamiltonian that favors configurations
approximating a ``canonical cell tiling''~\cite{Hen91CCT,Mi93} (CCT). 
Our motives are (1) to infer the phason elastic constants of the CCT ,
which is difficult to simulate directly, and (2) 
to observe the crossover with decreasing temperature $T$
as clusters of small (rhombohedral) tiles get bound together into 
larger (CCT) ``supertiles''
in a toy model.
The same techniques will be essential in computing the phason elastic 
constants for models of specific real quasicrystals.

\section{Canonical-cell tiling and Hamiltonian}
Atomic structure models of $i$-AlZnMg~\cite{Wind94} [or $i$-AlMnSi~\cite{Mi96a}], 
and many related alloys, 
may be built by placing icosahedral ``Bergman'' [or ``MI''] clusters
on the nodes of networks with inter-node linkages of two lengths:
$b\equiv 2.75\aR$ along twofold symmetry directions or $c\equiv 2.38\aR$ along
threefold directions, which we shall call a ``$bc$-network.''
(The quasilattice constant $\aR$ is set to unity in this paper;
$\aR \approx 0.5$nm in real quasicrystals.)
A ``canonical-cell'' tiling~\cite{Hen91CCT} (CCT) is a special
$bc$-network built from
the four smallest polyhedral cells having $b$ or $c$ edges.

A Monte Carlo simulation of the CCT is practically intractable, 
since {\it local} tile reshufflings do not exist:
spatially extended clusters must be rearranged to reach another
valid tiling~\cite{Hen91CCT,Ox9397}.  
To circumvent this problem, our model system is a $bc$ network in 
which the CCT constraints are not imposed,  but are favored by a Hamiltonian 
and thus satisfied in the limit $T\to 0$. 
We emphasize that unlike matching-rule Hamiltonians,~\cite{steinhardt}
eq.~(\ref{eq-Ham12})  does {\it not} force a unique
quasiperiodic $T=0$ state; the ground state ensemble is still
a random tiling, but a different one (with less entropy) than at $T=\infty$.
Our configuration space 
is the random ``three-dimensional Penrose tiling'' (3DPT), 
which includes all packings of the well-known oblate and prolate
rhombohedra having edges of length $\aR\equiv 1$ along 5-fold symmetry directions. 
Similar simulations of another imperfect approximation of the CCT by
a $bc$-network, namely the lattice-gas 
of Ref.~\onlinecite{Mi93}, will be reported elsewhere.~\cite{LG} 

The 3DPT contains a special subset of vertices, called ``12-fold'' 
nodes, at which rhombohedron edges emanate
in all 12 of the five-fold symmetry directions~\cite{Hen86}.   
This 12-fold network approximates the CCT, since
the separation of nearby nodes is usually a $b$ or $c$ linkage;
however a ``short'' separation of $\aR$ (along 5-fold axes) 
occurs rarely.  For our simulations, we adopted the Hamiltonian 
   \begin{equation}
   {\cal H}_ {12} = - N_ {12}+\asb N_{sb}, 
   \label{eq-Ham12}
   \end{equation}
where $N_ {12}$
is the total number of 12-fold nodes, 
and $N_{sb}$ is the number of ``short'' node pairs with
separation  $\aR$. 

The purpose of $\asb$ is to suppress ``short'' pairs;
$\asb\equiv 1$ suffices, which we adopted after testing other values.
We conjecture (and simulations support) that
the ground states of (\ref{eq-Ham12}) are a subset of CCT's, 
namely those with maximum node density, 
and not the maximum-entropy random CCT. 

\section {Perpendicular space and fluctuations}
Our rhombohedron edges are $\pm \epar_\alpha$, where~\cite{JarNel88}
     $\epar_1= (\tau,0,1)/(1+\tau^2)^{1/2}$
and the other basis vectors are made by 
reflections in the $x$, $y$, or $z$ planes, or by
cyclic permutation of the components. 
Thus every vertex has coordinates of the form
\begin{equation}
   \xpar = \sum_{\alpha=1}^6 
   n_\alpha \epar_\alpha,
   \label{eq-xpar}
\end{equation}
In a standard trick, the integer coefficients may be visualized as 
coordinates $\xx = [n_1, \ldots, n_6]$ of a 6-dimensional hypercubic lattice,  
into which eq.~(\ref{eq-xpar}) 
embeds a 3-surface  \cite{El85}. 
This 3-surface is conveniently parametrized by a 
``perpendicular'' (``perp'')
coordinate, defined for each vertex by
\begin{equation}
   \xperp= \sum_{\alpha=1}^6 n_\alpha \eperp_\alpha, 
   \label{eq-xperp}
\end{equation}
where $\eperp_1=\eta (1,0,-\tau)/(1+\tau^2)^{1/2}$
and the other basis vectors 
$\eperp_\alpha$ are derived by the same reflections and
cyclic permutations used to produce 
$\epar_ \alpha$ from $\epar_1 $.
We write $\hh(\rr)$ for the 
smoothed (coarse-grained) version of $\xperp$. 

The gradients $\nabla \hh(\rr)$, forming the ``phason strain''
tensor, quantify the local deviation from icosahedral symmetry.
The random-tiling scenario~\cite{Hen91ART} predicts that
a quasicrystal's free energy density has
a {\it phason elastic} form, 
\cite{Elastic}
i.e. is quadratic in components of $\nabla \hh$.
This was confirmed numerically for several completely random
($T=\infty$) tilings \cite{Hen91CCT,Ox9397,2DTM,Tang90,Shaw91a}, 
but not until now for the CCT. 
In contrast, the ``ideal tiling'' behavior implies
a free energy scaling as $|\nabla \hh|$.
An icosahedral quasicrystal has two elastic constants, 
$K_1$ and $K_2$.
Very roughly speaking, $K_1$ parametrizes the mean strength
of the elastic stiffness, and $K_2$ parametrizes its 
anisotropy (degree of coupling between direction of the
gradient in real space and components of $\hh$).

For fluctuations around a state with 
zero mean phason strain, the dimensionless elastic free energy 
is~\cite{Elastic,JarNel88}
  \begin{equation}
      F/T  = {1\over 2} \sum _{\qq,ij}
           \CK_{ij}(\qq) \tilde{h}_i(\qq) \tilde{h}_j(-\qq)
  \label{eq-Ffourier}
  \end{equation}
wehere it is convenient to use 
$\thh(\qq) \equiv V^{-1/2} \int d^3 \rr e^{-i\qq\cdot\rr} \hh(\rr)$.  
Here
  \begin{eqnarray}
    \CK_{ij}(\qq)& = & K_1|\qq|^2\delta_{ij}-
         K_2\; \big[({1\over 3}|\qq|^2+2q_i^2 \nonumber \\
    && + q_{i+1}^2/\tau-\tau q_{i-1}^2)\; \delta_{ij}-2q_iq_j \big].
  \label{eq-kquad}
  \end{eqnarray}
Eq.~(\ref{eq-Ffourier}) implies 
that each $\thh(\qq)$ consists of Gaussian random variables, and
their correlations 
  \begin{equation}
         \Corr_{ij}(\qq) \equiv 
          \langle {\tilde h}_i(-\qq)  {\tilde h}_j(\qq)\rangle. 
    \label{eq-Corr}
  \end{equation}
are given by
  \begin{equation}
         \Corr_{ij}(\qq) = [\CK(\qq)^{-1}]_{ij} .
  \label{eq-flucts}
  \end{equation}
The diffuse diffraction intensity~\cite{diffuse91}
near a Bragg peak with reciprocal lattice
vector $\Gpar$ is 
    \begin{equation}
      I(\qq) \propto \sum_{ij} G^\perp_i \Corr_{ij}(\qq) G^\perp_j , 
    \label{eq-diffuse}
  \end{equation}
where $\Gperp$ is the perp-space partner of $\Gpar$. 
(An equilibrium  random tiling 
in $d>2$ has true Bragg peaks~\cite{El85,Hen91ART} as
well as diffuse scattering.)
Experiments measure (\ref{eq-diffuse});
simulations can measure  (\ref{eq-Corr}) directly. 

\section {Simulation, measurements, and corrections}
Our data is taken from tilings using periodic boundary conditions
in a cubic cell of side
$L=\tau^6 (2-2/\sqrt5)^{1/2}\aR=18.868\aR$.
This corresponds to the ``8/5 Fibonacci approximant'' 
of the 3DPT~\cite{Shaw91a}, 
and contains $N_v =10 336$ rhombohedron vertices. 
We use the very simple update move for the 3DPT 
which rearranges the four tiles surrounding a vertex
with just four edges.\cite{Tang90,Shaw91a}.

Simulation time is measured in Monte Carlo sweeps, 
where each sweep makes $N_v$ actual flips on
randomly chosen vertices.
A simulation run consisted of $500$ thermal cycles;
each cycle included 3000 MCS at $T=\infty$  (to erase 
memory of the preceding cycle), 
followed by 34 cooling stages each at a constant temperature. 
Each stage takes 1000 MCS, except the first six stages ($T>0.8$)
used 200 to 800 MCS; 10 measurements were made during the last 100 MCS of each stage.
A typical 8/5 cooling run took 10 days of CPU time on a 1.5 GHz 
Athlon processor. 

We constructed $\xpar$ and $\xperp$ for each vertex, using
(\ref{eq-xpar}) and (\ref{eq-xperp}).
For each $\bf q$ on a standard list of 70 wavevectors, we took
   \begin{equation}
   \thvert(\qq)\equiv 
        {\sqrt{V}\over {N_v}}
        \sum_{\xx \in {\rm 3DPT}}\xperp e^{-i\qq\cdot \xpar}
   \label{eq-thvert}
   \end{equation}
as in Ref.~\onlinecite{Shaw91a},
where $V$ is the simulation cell volume and the
sum runs over the $N_v$ actual 3DPT vertices. 
The normalization in (\ref{eq-thvert}) ensures that
$\thvert(\qq) \approx \thh(\qq)$ for small $\qq$.
Fluctuation correlations analogous to (\ref{eq-Corr}), 
$\CorrH_{ij}(\qq)\equiv \langle \thvert_i(-\qq)\thvert_j(\qq)\rangle$
were accumulated, and were symmetrized with respect to 
cyclic permutations $(xyz)$. 

Although $\langle \thh(\qq) \rangle =0$, 
Eq.~(\ref{eq-thvert}) has a nonzero 
expectation when $\qq$ is a Bragg vector, 
just like the structure factor, from which 
the sum (\ref{eq-thvert})
differs only by the extra factor $\xperp$.
As in the structure factor, each Bragg peak of our $\CorrH(\qq)$ function
is surrounded by a tail of diffuse scattering, which must also be
subtracted in order to correct the nearby wavevectors. 
To calculate this tail, we adopt the ``undulating cut approximation''
(Ref.~\onlinecite{Hen91ART}, Sec. 7),  
in which the occupation probability of
a physical-space vertex $\xpar$ is $\rhoo(\xperp-\hh(\xpar))\leq 1$.
Here $\xpar$ and $\xperp$ are related as in eqs. (\ref{eq-xpar})
and (\ref{eq-xperp}), 
$\hh(\rr)$ is smooth and fluctuating according to (\ref{eq-flucts}), 
and $\rhoo(\hh)$ defines an ideal reference structure 
when $\hh$ is flat. 
The measured fluctuation is predicted to be~\cite{LG}
   \begin{equation}
       \CorrH_{ij}(\qq) = 
       \Corr_{ij}(\qq) + \sum _{\Gpar} \CorrBragg{ij}(\qq,\Gpar).
   \label{eq-sumBragg}
   \end{equation}
where
    \begin{equation}
          \CorrBragg{ij} (\qq;\Gpar) 
          =  D_i D_j \left[ V \delta_{\qq, \Gperp} +
              (\Gperp, \Corr (\qq-\Gpar) \Gperp) \right]. 
    \label{eq-Braggtail}
    \end{equation}
Here $\DD = \DD(\Gperp)$ can be expressed~\cite{LG} in terms of 
the Fourier transform of $\rhoo(\xperp)$;
in light of the symmetry, $\DD(\Gperp) \parallel \Gperp$ 
along any symmetry axis. 

Our measurements include
one orbit of weak Bragg peaks on 3-fold symmetry axes, 
-- at $\Gpar={{2\pi}\over L}(333)$ and ${{2\pi}\over L}(520)$ 
in our cell --
where the Bragg intensity contributes far stronger than the 
long-wavelength fluctuations (\ref{eq-Corr}). 
We determined each Bragg term (first term  in (\ref{eq-Braggtail}), 
i.e. $D_l D_m$) from
$\langle \thh(\qq)\rangle$ averaged 
over each temperature stage~\cite{rtdiff}.
>From our estimate of $K_1$ and $K_2$, we know $\Corr(\qq)$
via (\ref{eq-flucts}), hence the
second term in (\ref{eq-Braggtail}), so we can subtract the
sum in (\ref{eq-sumBragg}) from the measured data. 
The resulting 
matrices $\Corr(\qq)$ are inverted to obtain (via (\ref{eq-flucts})) 
estimates of (\ref{eq-kquad}), which are fitted to $K_1$ and $K_2$
using linear least squares.  The process is iterated until
converged.

\section{Results and Discussion}
\begin{figure}
\includegraphics[width=3.1in,angle=0]{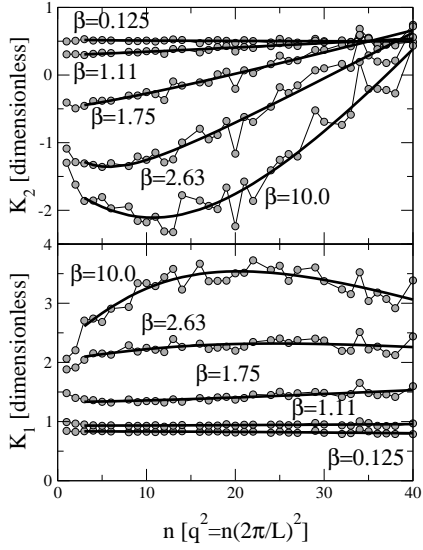}
\caption{
Phason elastic constants $K_1$ (bottom) and $K_2/K_1$ (top), where
$K_1$ and $K_2$ are fitted from
``perp'' space fluctuation $\CorrH_{lm}(\qq)$ of rhombohedron vertices, 
Fourier-transformed and separately fitted in each shell
of wavevector $\qq = \sqrt n (2\pi/L)$.  
Bragg peaks and tails were subtracted using (\protect\ref{eq-sumBragg}). 
Only five of the 34 temperatures are shown (labeled by
$\beta\equiv 1/T$. 
The fits to Eq.~(\ref{eq-fitform}) are shown as solid lines.
}
\label{fig-shells}
\end{figure}

To clarify two other artifacts which depend on $|\qq|$, we fitted
effective constants 
``$K_1(n)$'' and ``$K_2(n)$'' separately for each shell $n$ defined by
$|\qq|^2 = n (2\pi/L)^2$; the results are shown in Fig.~\ref{fig-shells}, 
for selected temperatures.  
The continuum approximation must break down 
at the CCT linkage length scale ($\sim 2.5\aR$), which explains the
deviations visible for $n\ge20$, i.e. $2\pi/|\qq|\leq 4.2 \aR$.
(Above the supertile formation temperature, the tile length
scale is the rhombohedron edge $\aR$, and no such deviations 
are seen in this wavevector range.)

At small wavevectors, equilibration time becomes problematic.
As shown in Ref.~\onlinecite{Shaw91a}, eqs. (20ff), 
the relaxation rates of $\thh(\qq)$ are $\gamma \CK_i(\qq)$, 
where $\gamma(T)$ is a relaxation coefficient and 
$\{\CK_1, \CK_2, \CK_3 \}$ are the eigenvalues of $\CCK(\qq)$. 
This becomes problematic when temperature 
is low (since $\gamma(T) \to 0$ as $T\to 0$)
and $|\qq|$ is small (as $\CCK \propto |\qq|^2$, 
eq.~(\ref{eq-kquad}). The consequence is that, at the
lowest temperatures in Fig.~\ref{fig-shells}, 
the smallest wavevectors 
are governed by the elastic constants 
of the higher temperature at which 
these modes fell out of equilibrium. 

The plots in Fig.~\ref{fig-shells} suggest two regimes 
as a function of wavevector and temperature.~\cite{FN-n20anomaly}
At high $T$ or large wavevector, 
a linear dependence on $n\propto |\qq|^2$ is visible, 
with the slope increasing as $T\to 0$. 
We attribute this to our use of the 3DPT vertices
in (\ref{eq-thvert}): the $\xperp$ differences 
between adjacent vertices (separated by $\aR$)
certainly exceed the fluctuations of the smooth $\hh(\rr)$ 
surface, extrapolated to this short scale.  Consequently, the 
inferred elastic constant is spuriously small. 

On the other hand, a second regime is visible 
at low $T$ and $|\qq|< \kappa(T)$, 
where $\kappa(T)$ defines a crossover. 
In this regime, $K_i(n)$ has a clear linear dependence, 
which we tentatively attribute to higher-order-gradient
terms of $O(q^4)$ 
in the harmonic phason-elastic coefficients (\ref{eq-kquad}), 
which are permitted by symmetry. 
Although it certainly affects the $n=1,2$
wavevector shells at the lowest temperatures,
lack of equilibration cannot explain all of the low $T$/small $|\qq|$ 
regime, as the data (in this regime) agree very well with 
our lattice-gas simulations~\cite{LG} which have a different dynamics.
We used the empirical fitting form 
  \begin{eqnarray}
   K_i(n) = a_i + b_i\sqrt{n^2+\nstar^2}+ c_i
        \frac{n}{\sqrt{n^2+\nstar^2}}
  \label{eq-fitform}
  \end{eqnarray}
to extrapolate $K_1(n)$ and $K_2(n)$ to $\qq=0$.

\begin{figure}
\includegraphics[width=2.8in,angle=-90]{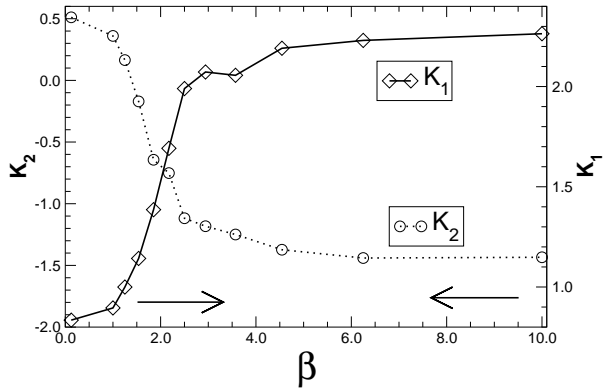}
\caption{
Elastic constants as a function of inverse temperature $\beta$. 
These are same fitted values indicated by solid lines
in Fig.~\ref{fig-shells},
extracted from the perp-coordinates of rhombohedron vertices
and extrapolated to $q\to0$.}
\label{fig-K1K2}
\end{figure}

Our principal result is the 
temperature dependence of the elastic constants indicated on 
Fig.~\ref{fig-K1K2}.
At infinite temperature, i.e. for the maximally random tiling, 
we fit $K_1= 0.84$ and $K_2=0.52$, so 
$K_2/K_1=0.62$ 
in agreement with previous measurements 
on the same ensemble~\cite{Tang90,Shaw91a}. 
As temperature decreases, $K_1$ increases, 
as expected since the Hamiltonian (\ref{eq-Ham12}) favors 
formation of ``supertiles'' %
which have small perp-space differences
among their corner nodes.
Meanwhile, $K_2/K_1$ turns negative with decreasing temperature, 
consistent with our knowledge that 
the random CCT\cite{New95b}  has $K_2<0$. 
It appears that $K_1 \to 2.26$ and $K_2 \to -1.44$ as $T\to 0$, 
thus $K_2/K_1 \to -0.64$, with errors of order 10\%. 
Our lattice-gas simulations\cite{LG}
gave similar results.

Approaching $T=0$, the fit of our data to the angular dependence 
of elastic theory may be worsening.  Furthermore, 
$K_2/K_1$ seems to be approaching the critical 
value $-0.75$, at which the system goes unstable to a 
modulation with $\qq$ and $\thh(\qq)$ along a 3-fold axis.
This suggests that the true canonical-cell tiling -- more precisely
the maximum-density subensemble of the CCT -- might have a more
complicated dependence than elastic theory. This is not 
unprecedented: exactly that is known to happen in those 
10-fold and 12-fold symmetric two-dimensional tilings~\cite{Ox9397}
that have extra constraints like the CCT. 

Neutron and X-ray diffraction measurements of diffuse wings around
Bragg peaks yield $K_2/K_1=-0.5$ for $i$-AlPdMn~\cite{Bois95,Le01,Le99}, 
agreeing with our result, so $i$-AlPdMn might
be modeled by a CCT-like network of microscopic clusters. 
On the other hand, $i$-AlCuFe has $K_2/K_1>0$~\cite{Le99}.
Thus, despite their similar face-centered icosahedral
atomic structures, $i$-AlPdMn and $i$-AlCuFe must be modeled by
different tile Hamiltonians.

The recent experimental calibration of $i$-AlPdMn data~\cite{Le01}
yielded $K_1/T=0.002 \Ao^{-3}$ in absolute physical units.
Multiplying by $\aR^3$ (with $\aR=4.5 \Ao$) converts this to 
$K_1 \approx 0.2$ in the units of this paper, which is puzzlingly small. 
This is hard to explain by a misjudgment in our model formulation.
If our tiles are invalid, one expects that 
(since this alloy is so well ordered), 
they should be replaced by  large supertiles, 
so that $K_1$ becomes {\it larger}, or that matching rules
are satisfied implying $K_1 \to \infty$. 
Instead, the data suggest that, implausibly, atoms are free to fluctuate 
in groups even {\it smaller} than a rhombohedron.

The temperature dependence of $K_i(T)$ was
used for extrapolation  only;
the Hamiltonian (\ref{eq-Ham12}) 
simply implements the CCT and is unlikely to model 
a realistic quasicrystal.
A physically relevant model would add to (\ref{eq-Ham12})
an additional perturbation $\alpha'$ that breaks the degeneracy 
among canonical-cell tilings, 
so that large $\alpha'$ corresponds to physical low temperature. 
Such a model can be investigated using the techniques of this paper. 

In conclusion, we evaluated the phason elastic constants of 
the canonical-cell tiling for the first time, by a 
simulation in which the mean-square fluctuations 
of the abstract surface representing the tile configuration
were measured at many wavevectors.
This is the first theoretical measurement in any
quasicrystal model of the temperature dependence 
of the phason elastic constants.

\begin{acknowledgments}
This work is supported by DOE grant DE-FG02-89ER45405;
computer facilities were provided by the Cornell Center for Materials
Research under NSF grant DMR-9632275. 
We thank M. de Boissieu for discussions. 
\end{acknowledgments}


\begin{thebibliography}{99}

\bibitem [*]{MM-addr} 
Materials Research and Liquids, Institute of Physics,
TU Chemnitz, D-09107 Chemnitz, Germany;
Current and permanent address, 
Institute of Physics, Slovak Academy of Sciences, 84228 Bratislava, Slovakia.

\bibitem{niceqxtal} 
See {\it e.g.} S.~W.~Kycia {\it et al}, 
Phys. Rev. B 48, 3544 (1993). 

\bibitem{Hen91ART}
C.~L.~Henley,
``Random tiling models,'' p. 429 in {\it Quasicrystals: The
State of the Art}, ed. P.~J.~Steinhardt and D.~P.~DiVincenzo,
(World Scientific, 1991).

\bibitem{Hen91CCT}
C. L. Henley, Phys. Rev. {\bf B43}, 993 (1991).

\bibitem{Wind94}
M.~Windisch, J.~Hafner, M.~Kraj\v{c}i and 
M.~Mihalkovi\v{c}, Phys. Rev. B 49, 8701 (1994).

\bibitem{Mi96a}
M. Mihalkovi\v{c} {\it et al}, 
Phys. Rev. B 53, 9002 (1996).

\bibitem{Matching}
K.~Ingersent in {\it Quasicrystals: The State of the Art}, 
eds. D. P. DiVincenzo and 
P.~J.~Steinhardt (World Scientific, 1991). 

\bibitem{Jeong94} 
H.-C. Jeong and P.J. Steinhardt, {Phys. Rev. Lett.} {73}, 1943 (1994).

\bibitem{diffuse91}
M. Widom, Phil. Mag. Lett. 64, 297 (1991);
 Y. Ishii, Phys. Rev. B45, 5228 (1992).


\bibitem{Bois95}
M.  de Boissieu, {\it et al},
Phys. Rev. Lett. 75, 89 (1995).

\bibitem{Le01}
A. L\'etoublon {\it et al}, 
Phil. Mag. Lett.  81, 273 (2001). 


\bibitem{Le99}
A.~L\'etoublon, thesis (Grenoble, France), 1998.

\bibitem{Mi93} 
M. Mihalkovi\v{c} and P. Mrafko, Europhys. Lett. 21, 463 (1993).



\bibitem{Ox9397}
M. Oxborrow and C. L. Henley, 
Phys. Rev. B 48, 6966 (1993);
M. Oxborrow and M. Mihalkovi\v{c},
p.~451 in
{\it Aperiodic '97}, 
eds. M.~de Boissieu, J.-L. Verger-Gaugry, and R.~Currat
(World Scientific, 1998).

\bibitem{steinhardt}
H.-C. Jeong and P.J. Steinhardt {Phys. Rev. B} {48}, 9394 (1993);
T. Dotera and P. J. Steinhardt, 
Phys. Rev. Lett.  72, 1670 (1994). 


\bibitem{LG}
M.~Mihalkovi\v{c} and C.~L.~Henley, in preparation.

\bibitem{Hen86}
C. L. Henley, 
Phys. Rev. B 34, 797 (1986).


\bibitem{JarNel88} 
M.~V.~Jari\'c and D.~R.~Nelson, Phys. Rev. B 37, 4458 (1988).

\bibitem{El85} 
V. Elser, Phys. Rev. B {\bf 32}, 4892 (1985).

\bibitem{Elastic}
T. C. Lubensky, in
{\it Aperiodic Crystals I: Introduction to Quasicrystals},
M. V. Jari\'c (ed.), (Academic Press, London, 1988),
and references therein.




\bibitem{2DTM} 
  W. Li, {\it et al}, J. Stat. Phys. {\bf 66}, 1 (1992), and
references therein.

\bibitem{Tang90}  
L.-H. Tang, Phys. Rev. Lett. {\bf 64}, 2390 (1990).

\bibitem{Shaw91a}
L. J. Shaw, V. Elser, and C. L. Henley, 
Phys. Rev. B43, 3423 (1991). 


\bibitem{rtdiff}
C.~L.~Henley, V.~Elser, and M.~Mihalkovi\v{c}, 
Z. Kristallogr 215, 553 (2000).

\bibitem{FN-n20anomaly}
We do not know the origin of the anomalies in the range
$12 \leq n \leq 20$ (the irregularities in
Fig.~\ref{fig-shells} are not statistical, 
except perhaps in the small-$|\qq|$/low-$T$ limit.)



\bibitem{New95b}
M.~E.~J.~Newman and C.~L.~Henley,
Phys. Rev.  B52, 6386 (1995).





\end {thebibliography}
\end{document}